\documentclass[twocolumn]{aastex6}

\usepackage{natbib}
\usepackage{amsmath}
\usepackage{times}
\usepackage{graphicx}
\usepackage{epstopdf}
\usepackage{textcomp}
\usepackage{color}

\newcommand{\myemail}{eva.wirstrom@chalmers.se}
\newcommand{\Ox}{O$_2$}
\def\cm3{cm$^{-3}$~}

\def\s1{s$^{-1}$}

\def\ltsim{\lower 0.7ex\hbox{$\buildrel < \over \sim\$}}  
\def\gtsim{\lower 0.7ex\hbox{$\buildrel > \over \sim\ $}}

\newcommand{\kms}{km s$^{-1}$}

\shorttitle{Search for Interstellar O$_2$}
\shortauthors{Wirstr\"om et al.}

\begin{document}

\title{ A Search for O$_2$ in CO-depleted Molecular Cloud Cores with {\it Herschel} }

  \author{ Eva  S. Wirstr\"om\altaffilmark{1} }
\affil{Department of Earth and Space Sciences, Chalmers University of Technology, Onsala Space Observatory,  439 92 Onsala, Sweden} 

\author{Steven B. Charnley and Martin A. Cordiner\altaffilmark{2}}
\affil{Astrochemistry Laboratory,  Mailstop 691, NASA Goddard Space
  Flight Center, 8800 Greenbelt Road, Greenbelt, MD 20770, USA}
    
\and 
\author{Cecilia Ceccarelli}
\affil{$^4$Laboratoire d'Astrophysique, Observatoire de Grenoble, BP 53, 38041 Grenoble cedex
09, France}

\altaffiltext{1}{\myemail}
\altaffiltext{2}{Department of Physics, The Catholic University of America, Washington, DC 20064, USA}

\begin{abstract}

The general lack of molecular oxygen in molecular clouds is an outstanding problem in astrochemistry.
 Extensive searches with {\it SWAS}, {\it Odin} and  {\it Herschel} have only produced two detections;  upper limits to the O$_2$ abundance in the remaining sources observed are about 1000 times lower  than predicted by chemical models. 
 
Previous atomic oxygen observations  and inferences from  observations of other molecules indicated that high abundances of O atoms might be present in dense cores exhibiting large amounts of CO depletion. Theoretical arguments concerning the oxygen gas-grain interaction  in cold dense cores suggested that, if O atoms could survive in the gas after most of the rest of the heavy molecular material has frozen out on to dust, then O$_2$ could be formed efficiently in the gas.
Using {\it Herschel} HIFI we searched  a small sample of four  depletion cores - L1544, L694-2, L429, Oph D -  for emission in the low excitation O$_2$ N$_J$=3$_3$--1$_2$ line at 487.249~GHz. Molecular oxygen was not detected and we derive upper limits to its abundance in the range 
 N(O$_2$)/N(H$_2$)$\approx (0.6 - 1.6) \times 10^{-7}$. We discuss the absence of O$_2$ in the light of recent  laboratory and observational studies.

\end{abstract}

\keywords{ISM: clouds - ISM: abundances - ISM: molecules - molecular
processes - star formation:chemistry }

\section{Introduction} \label{SectIntro}

Molecular oxygen has proven to be the most elusive molecule in the interstellar medium. Gas-phase chemical models predict that dense molecular gas (with hydrogen number densities $\sim$10$^3$--10$^5\rm~cm^{-3}$) should
contain large fractional abundances of O$_2$ ($\sim$10$^{-6}$--10$^{-5}$) after $\sim$10$^5$ years \citep[e.g.][]{Leung84,Bergin00}. However, extensive searches in various astronomical environments by {\it Herschel}\footnote{Herschel is an ESA space observatory with science instruments provided by European-led Principal Investigator consortia and with important participation from NASA.},  the {\it Submillimeter Astronomical Satellite (SWAS)} and {\it Odin}  (with respective beam sizes of $<$1\arcmin,  4\arcmin and 9\arcmin),  have demonstrated that  O$_2$ emission is extremely rare in molecular clouds. {\it SWAS} and {\it Odin} observations of molecular clouds constrained the O$_2$ abundance limits to lie in the range $\approx$(0.8--7)$\times$10$^{-7}$ \citep{Goldsmith00,Pagani03}.   Tentative detections by {\it SWAS} \citep{Goldsmith02}, and from ground-based observations of the $^{16}$O$^{18}$O isotopologue \citep{Pagani93}, both remain unconfirmed \citep{Pagani03,Fuente93,Marechal97}. Upper limits on the abundance of solid O$_2$ tend to rule out the possibility that the missing O$_2$ is hidden in icy grain mantles \citep{Vandenbussche99}.

The {\it SWAS} and {\it Odin} limits on O$_2$ abundance are several hundred times less than predicted in chemical models.  {\it Herschel} observations have pushed these limits down even more, to lie in the range $\approx$(0.1--5)$\times$10$^{-8}$ for many sources \citep{Goldsmith11IAU}.  Thus far, O$_2$ has only been securely detected in  Orion \citep{Goldsmith11a} and $\rho$ Oph A \citep{Larsson07,Liseau12}.  

Several theoretical explanations have been advanced to explain the low  O$_2$ abundances \citep[as summarised by][]{RobertsHerbst02}, but calculations show that, most probably, the simple process of oxygen atom accretion and ice formation on cold dust grains is responsible \citep{Bergin00, Charnley01,Viti01,RobertsHerbst02}. These models show that, for cloud ages $\gtsim$10$^{5}$ years, once CO formation is complete, significant O \added{atom} depletion is necessary to prevent the O$_2$ abundance from violating the observed limits.  However, it is generally unknown if the \replaced{O abundance}{elemental abundance of oxygen} in the observed molecular clouds is consistent with this explanation.  Observations and theoretical models available at the time of the {\it Herschel} mission indicated that O atoms might be transiently much more abundant than CO (and C atoms) in some dense cores, and so produce large abundances of O$_2$ in them.   In this paper we report the results of a  {\it Herschel} HIFI search for O$_2$ in four cores that exhibit high degrees of CO depletion.  Molecular oxygen was not detected and we provide upper limits.

\section{Oxygen in Dense Molecular Clouds}

Our search for molecular oxygen was based on the state of observational and laboratory knowledge at the time of  the {\it Herschel} mission.
  
\subsection{Atomic Oxygen}
Molecular oxygen  is easy to form in both warm  and cold dense gas through the neutral-neutral process 
\begin{equation} 
{\rm  O   ~+~   OH ~  \longrightarrow  ~ O_2~+~  O  }
 \label{Eq_c+oh}
\end{equation}
and is rapidly destroyed in the chemical reaction with atomic carbon
\begin{equation} 
{\rm  C   ~+~   O_2 ~  \longrightarrow  ~ CO ~+~  O  }
 \label{Eq_c+o2}
\end{equation} 
 Once the conversion \added{of atomic C to CO} is almost complete ($\sim$10$^{5}$ years), the O$_2$ abundance rises rapidly on a short time-scale. Collisions of O atoms with cold dust, leading to the formation and retention of water ice,  can occur on a comparable time-scale and  so inhibit interstellar O$_2$ production. 
 However, there is no direct evidence for low \added{atomic} O abundances in cold clouds. In fact, large O/CO ratio\added{s} \replaced{$>$15 has}{have} been measured \replaced{towards}{in several sources, e.g. O/CO$>$15 towards} W49N \citep{Vastel00}; where it has been attributed to selective CO depletion on grains\added{; and O/CO$\sim$50 towards L1689N \citep{Caux99}}.  High fractional abundances \added{of atomic O}, $\approx$10$^{-4}$, have also been measured in the collapsing envelope of the low-mass binary system IRAS 16293-2422 \citep{Ceccarelli00}.

L1544 belongs to a family of cold starless cores  in which CO molecules have been depleted from the gas \citep{Bergin02,Bacmann02,Crapsi05,BerginTafalla07}. \citet{Caselli02} modelled the ionization of L1544, as determined from observations  of $\rm HCO^+$, $\rm N_2H^+$  and their deuterated isotopologues, and  concluded that a large fractional abundance of \added{atomic} O of $\approx$10$^{-4}$  must be present  to destroy  $\rm H_3^+$  there, despite CO being significantly depleted on to dust. \citet{Gupta09} detected $\rm C_6H^-$ and  $\rm C_6H$ in the CO-depleted region of  L1544. Subsequent modeling by \citet{CordinerCharnley12} determined that a similarly high gas-phase O \added{atom} abundance was required to destroy $\rm C_6H^-$ and so account for the observed anion/neutral abundance ratio in this core, as well as in others that also show CO depletion (e.g. L1512).  In L1544, and other dense cores, some carbon-bearing molecules (CN, HCN, HNC) are also present in the CO-depleted gas \citep{Hily-Blant10,Padovani11} but at abundances much less than found in normal dark cloud gas, supporting the idea that C atoms are not particularly abundant in the gas.

 In summary,  there is some observational evidence that  suggest oxygen atoms may {\it not} freeze out efficiently in some dense cores, even when CO and some other heavy molecules (e.g. CS) do.  This, and the fact that the \deleted{C} abundance \added{of atomic C} must also be significantly reduced, raises the possibility that CO-depleted cores could contain large abundances of O$_2$.

\subsection{Gas-grain Chemistry of O$_2$ }

Could O$_2$ form in CO-depletion cores and be detectable? Here we demonstrate that this is theoretically possible. 
 Models of O$_2$ suppression have simply assumed that gas phase O atoms are hydrogenated to form H$_2$O ices (and so removed) at the rates at which they collide and stick to grains \citep[e.g.][]{Bergin00, Viti01,Charnley01,RobertsHerbst02}; experiments have confirmed the viability of this  ice formation pathway \citep{Dulieu10}. 
However, to maintain a population of O atoms in gas where CO is freezing out requires that collisions of oxygen atoms with cold grains do not result in 100\% retention on the surface. \citet{Maret06} have shown that the CO-depletion core B68  contains a significant abundance of N atoms as well as N$_2$ molecules and so it is reasonable to inquire whether this could also occur for O atoms.

Assumptions have been made in previous chemical models that could allow the  return of O atoms to the gas at non-negligible rates \citep[e.g.][]{HasegawaHerbst93,WillacyWilliams93,Charnley01,Viti01,Caselli02, CharnleyRodgers02,Flower05,Crapsi05} and include: (i) lower than expected binding energies on the relevant surface, $E_{\rm 0}$\,=\,600 K, \citep[cf. 800 K,][]{TielensAllamandola87}, or (ii) a low sticking efficiency, or (iii) a variety of non-thermal desorption processes such as photodesorption or cosmic ray impacts \citep{Leger85}, or (iv) desorption induced by H$_2$ formation \citep{WillacyMillar98}. Thus, accepting that at least one of the above assumptions holds, it is feasible for an accreted O atom to desorb from a cold grain. However, since clearly {\it some} oxygen atoms must be retained on grains,  the gas-grain kinetics of  O atoms and CO molecules has to incorporate some degree of selective retention on the dust.

\citet{Hollenbach09} have presented a model for photodissociation regions (PDRs), in which the O atoms can be thermally desorbed, and which treats water formation in a more kinetically realistic manner. On the grain surface,  H atom addition first converts \added{atomic} O to OH, and these hydroxyl molecules can then form hydrogen bonds with pre-existing water molecules;  if an O atom does not meet a surface H atom there is a finite probability that it will desorb and return to the gas.  The strengths of hydrogen bonds are typically at least a factor of ten greater than those for van der Waals (physisorption) bonding, and it is this  {\it chemical fixing} which controls the rate at which  gaseous \added{atomic} O can become depleted on grains and suppress O$_2$ production.  Thus, O \added{atom} accretion onto grains by {\it fixing}, as opposed to simply {\it sticking} at the gas-grain collision rate, depends on the gas phase atomic O/H ratio, and is at its most efficient when the accreting gas has O/H$<$1,  and least so when oxygen atoms are more abundant than hydrogen atoms, i.e. when O/H$>$1.  

This naturally introduces a dependence on the density of the accreting gas.  Because O \added{atom} removal explicitly depends on the H atom kinetics, water ice formation can occur easily at low densities ($\sim$10$^4$~cm$^{-3}$ or less) where O/H$<$1 and the H/H$_2$ ratio has not yet reached a steady-state \citep[corresponding to $n(\rm H)\sim1\rm~cm^{-3}$,][]{GoldsmithLi05}. 
However, higher densities, $n$(H$_2$)$\sim10^6 \rm~cm^{-3}$,  lead to an increased  O/H ratio ($>$1), as H atoms are more efficiently converted to H$_2$ on grains, and consequently to less efficient O \added{atom} removal.  CO freeze-out in cold cores ($\sim$5--10~K) becomes apparent at  hydrogen densities in the range $n$(H$_2$)$\sim$1.5$\times$10$^4$--6$\times$10$^5$~cm$^{-3}$ \citep{Jorgensen05}, and so is concomitant with O/H$>$1. Thus, counterintuitively,  the central regions of dense cores exhibiting significant CO  depletion may contain large abundances of O$_2$ ($\sim$10$^{-6}$--10$^{-5}$). 

 An important point is that,  at about 10~K, this O$_2$ will not react with H$_3^+$ or any of its deuterated isotopologues, and so would not affect the observed ionization or  D/H enhancements found in these regions. The failure of  previous searches for O$_2$ towards starless cores could then be attributed to not studying cores of sufficiently high CO depletion \citep[L183 only has  $f_D$(CO)$\sim$5,][]{Pagani05}, misalignment of the telescope beam with the depletion peak \citep[$\sim$2\arcmin N of the TMC-2 position targeted by][]{Fuente93}, beam dilution \citep[{\it SWAS} and {\it Odin},][]{Goldsmith00, Pagani03}, or some combination of the above.

\added{One would expect that much of the H$_2$O formed from atomic oxygen will freeze out on to the dust, so will O$_2$ suffer the same fate?}
N$_2$ freeze-out appears to set in at densities of  $n$(H$_2$)$\sim$10$^6$~cm$^{-3}$ above that of CO \citep{BellocheAndre04,Pagani05,Pagani12}.  The cause of his selective depletion in cold cores is poorly understood \citep{BerginTafalla07} but is unlikely to be connected to  the difference in physisorption binding energies;  these are measured to be very close: 855$\pm25$~K for CO and 790$\pm25$~K for N$_2$ \citep{Oberg05}. Similarly, the  measured O$_2$ binding energy of 912$\pm15$~K  \citep{Acharyya07} is sufficiently close to those of  CO and N$_2$ that it is uncertain whether  O$_2$  could exhibit selective depletion  more like CO or  N$_2$ at high densities.   If the observed CO/N$_2$ depletion is  connected to the fact that these homonuclear diatomic molecules are nonpolar \citep{Masel96}, then it might be expected that O$_2$ will persist after CO has become depleted, up to the density where N$_2$ is depleted, and perhaps beyond.


\section{{\it Herschel} Observations}
In order to test whether O$_2$ abundances could be be significantly enhanced in high density depletion cores, we used the HIFI instrument \citep{deGraauw10} on-board the \textit{Herschel Space Observatory} \citep{Pilbratt10} to observe the N$_J$=3$_3$--1$_2$ O$_2$ line at 487.249~GHz towards a sample of such sources, listed in Table \ref{tab-sources}.
\begin{table}
\caption{Source list with observational parameters}
\begin{center}
\begin{tabular}{lcccc}
\noalign{\smallskip}
\hline\hline
\noalign{\smallskip}
   &RA (J2000.0) & Dec (J2000.0) & $V_{\rm LSR}$   & Obs Id    \\ 
 Source  &(~$^{\rm h}$~~$^{\rm m}$~~$^{\rm s}$~)  &(~$^{\circ}$~~\arcmin~~\arcsec~) & (\kms)  & \\  
\hline
 L429  & 18 17 05.1 &  -08 13 40 & 6.7  & 1342251645 \\ 
 Oph D  &  16 28 28.9 &  -24 19 19   & 3.5  & 1342250734 \\ 
 L1544  &   05 04 16.6 & +25 10 48  & 7.2  & 1342250746 \\ 
 L694-2  &   19 41 04.5  &  +10 57 02  & 9.6  & 1342245385 \\ 
\noalign{\smallskip}  
\hline
\end{tabular}
\end{center}
\label{tab-sources}
\end{table}
 The sources were  selected based on the known correlation between enhanced abundances of deuterated molecules and high CO depletion, and are the ones showing the highest CO depletion factors,  $f_D(\rm CO)$, in the surveys of H$_2$D$^+$, N$_2$H$^+$ and N$_2$D$^+$  in starless cores by \citet{Crapsi05} and \citet{Caselli08}.  Observations targeted the N$_2$D$^+$ emission peaks in the maps by \citet{Crapsi05}, which can be significantly offset from continuum core positions, but better tracing the depletion cores.  \deleted{The well-studied source L1544 was included in this search despite its low water content (Caselli et al. 2010).} Note that gas densities and temperatures in these cores are typically $n$(H$_2)>10^5$~cm$^{-3}$ and $T_K$$\sim$10~K, which makes the 487.249 GHz line ($E_u$=26~K) the most favourable one to observe in emission with HIFI, and collision rates fast enough for the fractional population of all energy levels to be well described by a Boltzmann distribution at the kinetic temperature.

The dual beam switching mode was employed with the LO frequency in HIFI band 1a set to 494.93 GHz in both L and R polarisation, placing the line in the lower side band. Both the Wide-Band Spectrometer (WBS) and the High Resolution Spectrometer (HRS) were used, and because emission lines from the region are known to be narrow (N$_2$D$^+$ FWHM $\sim$0.2-0.4 km\,s$^{-1}$) the HRS was configured to high resolution mode with a band width of 230 MHz, corresponding to a spectral resolution of 125 kHz, or $\sim$0.08~km\,s$^{-1}$ at this frequency.  Observations were performed under observing program OT2\_ewirst01\_1 and the data presented here is available from the \textit{Herschel} Science Archive\footnote{http://archives.esac.esa.int/hda/ui/} (HSA) under observing IDs given in Table~\ref{tab-sources}. 

The beam FWHM is 44$\arcsec$ at this frequency, the forward efficiency $\eta_{\rm l}$\,=\,96\%, and the main-beam efficiency $\eta_{\rm mb}$\,=\,76\%.  Detailed information about the HIFI calibration including beam efficiency, mixer sideband ratio, pointing, etc., can be found on the \textit{Herschel} web site\footnote{http://herschel.esac.esa.int/}. The in-flight performance is described by \citet{Roelfsema12}. 

Spectra of both polarisations were reduced separately using the \textit{Herschel} Interactive Processing Environment \citep{Ott10}. 
Subsequently, data FITS files were exported to the spectral analysis software XS\footnote{Developed by Per Bergman at Onsala Space Observatory, Sweden; http://www.chalmers.se/rss/oso-en/observations/data-reduction-software} for further reduction and analysis. After linear baseline subtraction and frequency alignment, the two polarisations for each observing ID and spectrometer were averaged together, weighted by rms noise. Pointing offsets between polarisations were within 7$\arcsec$, i.e., less than 20\% of the beam size.


\section{Results and analysis} \label{ResAnaSect}

 \begin{figure} 
\includegraphics[width=\columnwidth]{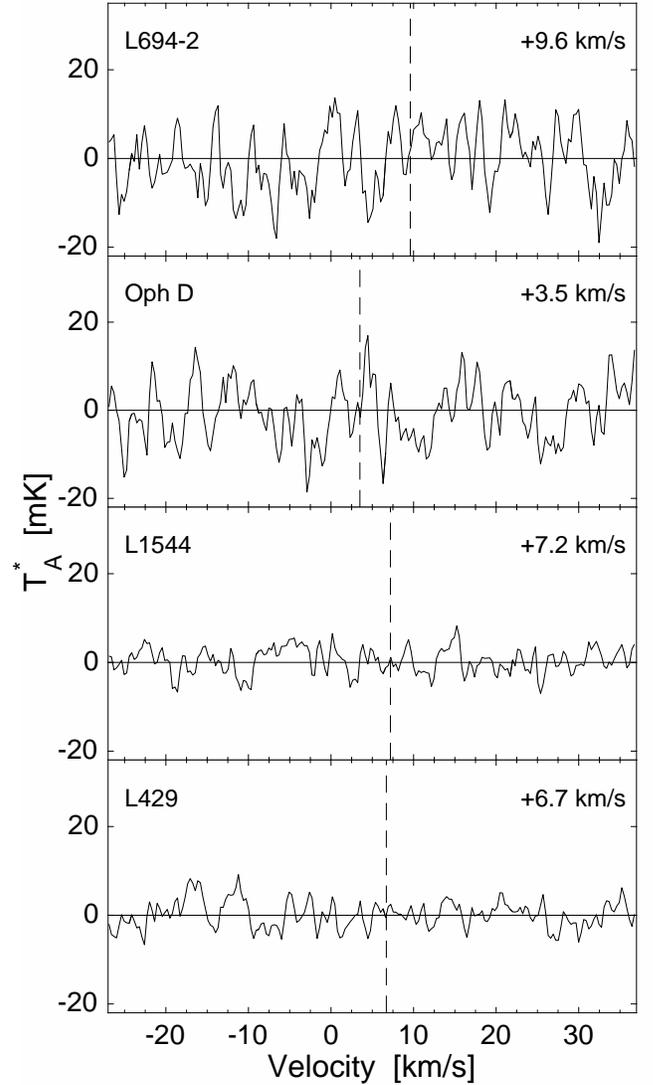} 
\caption{HIFI WBS spectra toward the four sources. The expected velocity of the \Ox\ 3$_{3}$ - 1$_{2}$ line, corresponding to the 
center velocity of N$_2$D$^+$ emission \citep{Crapsi05}, is given in the upper right corner of each spectrum and marked by vertical dashed lines to guide the eye. }
\label{fig-O2spectra}
\end{figure}
Figure \ref{fig-O2spectra} shows the observed Herschel HIFI Wide Band Spectrometer (WBS) polarization-averaged spectra towards the four sources in this study. Linear baselines have been subtracted and 
the resulting noise rms varies between 3--7 mK.

Molecular oxygen was not detected towards any of the cores. Before further analysis, spectra were multiplied by the total efficiency factor of $\eta_{\rm l}$/$\eta_{\mathrm{mb}}$=1.27 (HIFI Observers' Manual version 2.4) in order to get them on the $T_{\mathrm{mb}}$ scale. 
 For estimates of the \Ox\ abundance upper limits the non-LTE radiative transfer code RADEX was used to model the level populations of \Ox\ molecules and 3$\sigma$ line intensity upper limits. The spectroscopic data were adapted from \citet{Drouin10} and collision rates with H$_2$ scaled from He + \Ox\ collision rates of \citet{Lique10}. Physical conditions of the cores, listed in Table~\ref{TabSourceO2}, were adapted from \citet{Caselli08}, while line widths were taken to be the same as for N$_2$D$^+$. 
 \begin{table*}
\centering
\caption{Adopted source parameters and derived oxygen abundance limits}
\label{TabSourceO2}
\begin{center}
\begin{tabular}{lcccccccccc}
\noalign{\smallskip}
\hline
\hline
\noalign{\smallskip}
Source & $n$(H$_2$)$^a$ & $T_{\rm kin}^a$  & $f_D(\rm CO)^a$ & $\Delta v^b$ & $N$(H$_2$)$^a$ & $N$(O$_2$)$^c$ & $X$(O$_2$) & $f_{dil}$ & $X_{dil}$(O$_2$)$^d$ \\
 & cm$^{-3}$ & K & & \kms & cm$^{-2}$ & cm$^{-2}$ & & & \\
\noalign{\smallskip}
\hline
\noalign{\smallskip}
L429 & 6e5 & 7 & $>$20 & 0.4 &1.2e23 & $<$1.1e16 & $<$9.2e-8 & 0.25 &  $<$1.2e-6\\
Oph D & 5e5& 7 & $\sim$19  & 0.2 & 1.1e23 &$<$1.2e16 & $<$1.1e-7 & 0.32 &  $<$1.3e-6\\   
L1544 & 2e6 & 7 & $\sim$19 & 0.3 & 1.3e23 & $<$8.2e15 & $<$6.3e-8 & 0.17 &  $<$1.6e-6 \\
L694-2 & 9e5 & 7 & $\sim$16  & 0.3 & 1.1e23 & $<$1.8e16 & $<$1.6e-7 & 0.46 &  $<$1.3e-6 \\
\noalign{\smallskip}
\hline
\noalign{\smallskip}
\end{tabular}
\end{center}
\begin{flushleft}
 \footnotesize{$^a$From \citet{Caselli08}. $^b$From N$_2$D$^+$ lines \citep{Crapsi05}. $^c$Beam averaged, 3$\sigma$ upper limit. $^d$ Upper limit assuming the corresponding beam dilution factor $f_{dil}$, see Sect.~\ref{ResAnaSect} for more details.}
\noindent
  \\
\end{flushleft}
\label{tab-params}
\end{table*}
 Table~\ref{TabSourceO2} presents the beam averaged upper limits on O$_2$ column densities based on these calculations, and upper abundance limits to O$_2$/H$_2$ in the range $\approx$(0.6 -- 1.6)$\times$10$^{-7}$, assuming a constant H$_2$ column density over an area larger than the beam. 
 
 If the O$_2$ gas is only present in the inner, most CO-depleted parts of the cores, the emission would be significantly diluted in the large HIFI beam, and higher abundances could remain undetected. For example, assuming Gaussian O$_2$ source distributions with half intensities close to the 90\% contours of the N$_2$D$^+$ maps by \citet{Crapsi05}, the abundance limits are instead of the order of 10$^{-6}$, see Table~\ref{TabSourceO2} for corresponding beam dilution factors and abundances.

\section{Discussion}

 Recent experiments indicate a much higher physisorption binding energy for O atoms on amorphous water ice \citep[1660$\pm60$~K;][]{He15}. This is almost twice the value previously considered in chemical models \citep[800~K,][]{TielensAllamandola87} and was in fact suggested by \citet{Hollenbach09}.  Model calculations confirm that, in the case of PDRs, higher O atom sticking efficiencies lead to a significant  reduction in the gas phase O$_2$ abundance \citep{He15}. \added{Nevertheless, the low O$_2$ abundances inferred in CO depletion cores can only be explained by these models if the gas-phase atomic oxygen abundance is significantly lower than 10$^{-5}$.}
  
 The inferred low abundance of atomic oxygen in L1544 is consistent with the low gaseous H$_2$O abundance measured by \citet{Caselli12} since higher O atom abundances would also produce more water \citep[e.g.][]{Bergin00}. \added{On the other hand, it is too low to be consistent with the O atom abundances of $\leq$10$^{-4}$ required to explain the anion/neutral ratios in the same source \citep{CordinerCharnley12}.} Experiments show that O atoms could diffuse on grain surfaces faster than by pure thermal hopping \citep{Minissale14} and so, when O/H$>$1 in the accreting gas, reactions between O atoms \citep{TielensHagen82} could form solid O$_2$ efficiently \citep[cf.][]{Charnley05}, although some may  be hydrogenated to H$_2$O  \citep{Ioppolo08,Oba09}.   Photodesorption of this O$_2$ would be dissociative \citep{Fayolle13} and would not increase the  O$_2$ abundance in the gas of, say, L1544. \added{Also, O$_2$ is predicted to have a higher physisorption binding energy to amorphous water ice when considering a low fractional coverage \citep[1310 K;][]{He16} compared to previous experimental estimates considering monolayer coverage \citep{Acharyya07}. This effect is even larger for CO and N$_2$, resulting in an O$_2$ sticking efficiency significantly lower than that of CO, and following more closely that of N$_2$ \citep{He16}. }
However, competition for  binding sites between these O$_2$ molecules and CO molecules arriving on the surface  could lead to their ejection into the gas \citep{Noble15}.  
    Our O$_2$ non-detections could therefore place limits on these processes in CO-depleted cores.
    

  We did not detect molecular oxygen at abundances as high as $10^{-6}$ in the survey of depletion cores. The derived upper limits of $\sim$10$^{-7}$ or less are similar to those obtained elsewhere. \citet{Yildiz13} determined  O$_2$/H$_2 \leq 6 \times 10^{-9}$ in  NGC 1333 IRAS 4A although there is a  tentative O$_2$ detection in the surrounding cloud material.  
  A low value for O$_2$/H$_2$ of $5 \times 10^{-8}$ has been detected in  $\rho$ Oph A by \citet{Liseau12}.  The widespread distribution of D-enriched formaldehyde and the detection of H$_2$O$_2$ in $\rho$ Oph A  \citep{Bergman11a,Bergman11b}  both support  the idea that   the O$_2$ could be present following the evaporation of  ice mantles from dust grains.  A search for H$_2$O$_2$ in Orion was unsuccessful \citep{LiseauLarsson15} although the upper limits for  hydrogen peroxide in interstellar ices,  H$_2$O$_2$/H$_2$O$<9\%$, do not rule out it being quite abundant on dust grains \citep{Smith11}.
       In Orion, the O$_2$ emission is confined to the H$_2$ Peak 1 position and the originally reported abundance ratio was O$_2$/H$_2\sim10^{-6}$ \citep{Goldsmith11IAU}.     A combination of  postshock chemistry in a UV-illuminated  MHD shock wave with a fortuitous line-of-sight can account for the presence of O$_2$  in Orion, and suggest that the O$_2$ abundance could possibly be even higher $\sim$10$^{-5}$--10$^{-4}$ \citep{Chen14,MelnickKaufman15}.  In Sgr A,  \citet{Sandqvist15} find O$_2$/H$_2 \leq 5 \times 10^{-8}$ but also suggest that higher abundances, O$_2$/H$_2 \leq 1-2 \times 10^{-5}$,  could exist in foreground clouds.

   
  In conclusion, as with much of the dense interstellar medium, O$_2$ is underabundant in CO-depletion cores, most probably due to a large degree of O atom freezout to form water ice.  The reasons why it remains so generally underabundant and yet is detectable in a few sources remains a mystery\added{, and there is little prospect of detection in additional sources in the near future with Herschel concluded and no follow-up space-based spectrometer in the pipeline}.   Nevertheless, understanding the interstellar O$_2$ deficiency can shed light on many poorly understood issues in astrochemistry \citep{Melnick12}.

\acknowledgments

E.S.W. acknowledges generous financial support from the Swedish National Space Board. The work of  S.B.C. and M.A.C.  were supported by NASA's Origins of Solar Systems Program.

\vspace{5mm}
\facilities{Herschel(HIFI)}

\bibliography{references}

\end{document}